\documentclass[11pt]{article}
\usepackage{amsfonts}

\usepackage{graphicx}

 \def\be{\begin{equation}}
 \def\ee{\end{equation}}
 \def\bea{\begin{eqnarray}}
 \def\eea{\end{eqnarray}}

 \def\1{\'{\i}}

\newcommand{\te}{\phi}

\newcommand{\xx}{X}

\newcommand{\yy}{Y}

\newcommand{\mm}{{\cal E}}

\newcommand{\mpt}{{\cal E}}

\begin{document}

 \noindent
 \medskip

\begin{center}

{\sc{\Large{Factorization approach to superintegrable systems:\\[4pt]
Formalism and applications}}}

\end{center}

\medskip
 
\begin{center}
{\sc \'Angel Ballesteros$^a$,      Francisco J.~Herranz$^{a,}$\footnote{
  Based on the contribution presented at  ``The IX International Symposium on Quantum Theory and Symmetries" (QTS-9), July 13-18, 2015,  Yerevan,  Armenia.}, \c{S}engul Kuru$^b$ and Javier Negro$^c$}
\end{center}

\small
\noindent
{$^a$ Departamento de F\1sica,  Universidad de Burgos,
09001 Burgos, Spain\\ ~~E-mail: angelb@ubu.es,  fjherranz@ubu.es\\[10pt]
$^b$ Department of Physics, Faculty of Science,
 Ankara University, 06100 Ankara, Turkey\\
~~E-mail:  kuru@science.ankara.edu.tr\\[10pt]
$^c$ Departamento de F\'{\i}sica Te\'orica, At\'omica y
\'Optica, Universidad de Valladolid,\\  47011 Valladolid, Spain\\
~~E-mail:   jnegro@fta.uva.es

\normalsize
\medskip
\medskip

\begin{abstract}
The factorization technique for superintegrable Hamiltonian systems is revisited and applied  in order to obtain additional (higher-order) constants of the motion. In particular, the factorization approach to the classical anisotropic oscillator on the Euclidean plane is reviewed, and  new classical (super)integrable anisotropic oscillators on the sphere are constructed. The Tremblay--Turbiner--Winternitz system on the Euclidean plane is also studied from this viewpoint.
\end{abstract}
 
\medskip
\medskip
\medskip
\medskip
\medskip

\medskip


 \vfill
 \eject
 

\section{Introduction}

In some previous works \cite{Kuru3,Kuru4}, a new method to deal with the symmetries
of some superintegrable systems was introduced. This method  consists essentially
in extending the  factorization method for quantum mechanical Hamiltonians~\cite{infeldhull51} to some 
classical separable systems depending on several variables. We recall that if an integrable classical Hamiltonian $H$ can be separated in a certain coordinate
system, it is well-known that  each coordinate leads to an integral of the motion. Then, two sets of ``ladder'' $B^\pm$ and ``shift'' functions $A^\pm$ can be found and, if certain conditions are fulfilled, additional constants of motion can be explicitly constructed in a straightforward manner by combining these  $B^\pm$ and $A^\pm$ functions. It is worth stressing that such integrals are, in the general case, of higher-order on the momenta.

 This ``extended'' factorization method
has many advantages that we briefly enumerate:
i) The method is valid for quantum as well as for classical systems \cite{Kuru1}, and the
classical-quantum  correspondence becomes manifest at each stage of the factorization procedure. ii)~The approach can be applied either for
second-order or for higher-order symmetries. iii)~The symmetries so  obtained  
close a quite simple symmetry algebra \cite{Kuru4} from which it is straightforward
to write the associated polynomial symmetries. iv) For classical systems the results allow to find in a simple way the associated phase space trajectories, and in the case of quantum systems the discrete spectrum can be explicitly computed.

The aim of this work is to provide an introduction to this method by means of some (known and new) examples of two-dimensional  superintegrable systems,
where we have restricted ourselves to the classical framework in order to make the
presentation simpler. However, we stress that the very same procedure can also be  applied to quantum Hamiltonians, thus leading to the corresponding ladder and shift operators. 

The structure of the paper is as follows. In the next section we   revisit the anisotropic oscillator on the 
Euclidean plane. Section 3 is devoted to propose an anisotropic oscillator Hamiltonian
on the sphere, which is a completely new model. In section 4 we  
study the classical Tremblay--Turbiner--Winternitz (TTW) system on the Euclidean plane from the factorization point of view. Finally, some remarks and conclusions  
close the paper.

\section{Anisotropic oscillators on the Euclidean plane}

Let us consider the  anisotropic oscillator Hamiltonian with unit mass and   frequencies $ \omega_x$ and $ \omega_y$ on the 
 Euclidean plane  in Cartesian coordinates:
\be
{H} = \frac12(p_x^2 + p_y^2) + \frac{ 1}{2}( \omega_x^2  x^2 + \omega_y^2y^2) .
\label{aa}
\ee
Obviously, this Hamiltonian   is   integrable since it Poisson-commutes with the (quadratic in the momenta) integrals of motion 
$$
{I}_x=\frac12 p_x^2  + \frac{ 1}{2} \omega_x^2  x^2,\qquad     {I}_y=\frac12 p_y^2  + \frac{ 1}{2}  \omega_y^2  y^2,
\qquad {H} ={I}_x+{I}_y.
$$
 
It is well-known that for  commensurate frequencies $\omega_x: \omega_y$ the Hamiltonian (\ref{aa}) defines a superintegrable system~\cite{Jauch, Stefan, Winternitz}, endowed with an  ``additional" integral of motion. 
In the sequel we study the Hamiltonian  (\ref{aa}) by following a factorization approach~\cite{Kuru3,Kuru4, Kuru1}  (see also~\cite{david,Kuru2} and references therein) which is different from the  one applied in~\cite{Jauch, Stefan,  Winternitz}. 

Firstly, let us introduce a  positive real parameter $\gamma$:\be
 \omega_x=\gamma \omega_y ,\qquad  \omega_y=\omega ,
 \label{ab}
\ee
 which encodes the anisotropy. 
This leads us to define a new coordinate $\xi$   as
\be
    \xi=\gamma x,\qquad  p_\xi =p_x/\gamma .
 \label{ab2}
\ee
Hence the   Hamiltonian (\ref{aa})  is expressed  as
\bea
&& H = \frac12(p_x^2 + p_y^2) + \frac{\omega^2}{2} \left( (\gamma x)^2 + y^2 \right)  \nonumber\\[2pt]
&&\quad \,
= \frac12\, p_y^2 + \frac{\omega^2}{2} \,  y^2 
 +  \gamma^2\left(\frac12\, p_\xi^2 + \frac{\omega^2}{2\gamma^2} \,  \xi^2\right) .
 \label{ae}
\eea
In this latter form, the  two one-dimensional Hamiltonians $H^\xi$ and $H^y$
\be
 H^\xi = \frac12\, p_\xi^2 + \frac{\omega^2}{2\gamma^2} \, \xi^2  ,\quad\   H^y = \frac12\, p_y^2 + \frac{\omega^2}{2} \, y^2 ,
\label{af}
\ee
are two integrals of motion quadratic in the momenta, since $\{ H , H^\xi \}=  \{ H , H^y \}=  \{ H^\xi , H^y \}=0 $.
Therefore,  since the function $ H^\xi$ is a constant of motion, the complete Hamiltonian (\ref{ae}) is just  
$$H= H^y 
 +  \gamma^2 H^\xi ,
 $$
and $H$ is reduced to a one-dimensional system on the phase space submanifold defined by $H^\xi$ being a certain constant. 

As we will see in the sequel, the factorization approach requires, firstly,  to factorize the integral  $ H^\xi$ in terms of  ladder   functions $B^\pm$ and, secondly, to factorize   $H$ through $H^y$ in terms of shift functions $A^\pm$ (we remark that in this specific example 
the names ``ladder'' and ``shift'' can be interchanged). 

\subsection{Factorization}

Ladder  (lowering and raising)  functions $B^\pm$ for the constant of motion  $ H^\xi$ (\ref{af}) are obtained by imposing that 
\be
H^\xi =B^+ B^- +\lambda_{B} ,
\label{qa}
\ee
yielding
\begin{equation}
B^{\pm} ={\mp}\frac{i}{\sqrt{2}}\, p_{\xi}+
\frac{1}{\sqrt2} \frac{\omega}{\gamma}\,  \xi \, , \qquad \lambda_{B}=0 .
\label{bc0}
\end{equation}
The three functions $H^\xi$ and $B^\pm$ obey  the Poisson brackets given by
\be
\{H^\xi ,B^{\pm} \}=\mp i\, \frac{\omega}{\gamma} \,B^{\pm} , \qquad  \{ B^-,B^+ \}= - i\,\frac{\omega}{\gamma} ,
\label{commpt20} 
\ee
so that, together with the  constant function $1$, they  close the harmonic oscillator Poisson--Lie algebra $\mathfrak{h}_4$.

As far as the    shift functions $A^\pm$  is concerned, we factorize  $H^y$  (\ref{af})   by imposing that 
\be
H^y= A^+ A^-+\lambda_A ,
\label{qb}
\ee
giving rise to
\begin{equation}
A^{\pm}=\mp \frac{i}{\sqrt{2}}\,p_{y}-\frac{\omega}{\sqrt{2}}\,  y  ,\qquad \lambda_A=0  .
\label{capm0}
\end{equation}
Again,  the four functions  $(H^y,A^\pm,1)$   span the Poisson--Lie   algebra $\mathfrak{h}_4$ since
$$
\{H^y ,A^{\pm} \}=\pm  i  {\omega}  A^{\pm} , \qquad  \{ A^-,A^+ \}=   i {\omega}  .
$$

Thus, the two-dimensional Hamiltonian (\ref{ae}) can be expressed in terms of these ladder and shift functions as
$$
H= A^+ A^- +\gamma^2 B^+ B^-,
$$
and the following Poisson algebra is generated
$$
  \{H ,B^{\pm}\}=\mp i {\gamma} {\omega} B^{\pm}   ,\qquad \{H ,A^{\pm}\}=\pm{ i \omega} A^{\pm}\,.
$$

\subsection{Higher-order integrals of motion}

The remarkable result arises if we consider a rational value for the parameter $\gamma$, namely, 
\be
\gamma  = \frac{\omega_x}{\omega_y}=\frac{m}{n}   , \qquad m,n\in \mathbb N^\ast .
\label{ah}
\ee
In such a case   ``additional" integrals of motion $X^{\pm}$ can be constructed for the Hamiltonian $H$  (\ref{ae}) by combining the ladder (\ref{bc0}) and shift (\ref{capm0}) functions in the form
\begin{equation}
X^{\pm}=(B^{\pm})^n (A^{\pm})^{m}\, ,   \qquad  \{H,X^{\pm}\}= 0 .
\label{csymmet10}
\end{equation}

Notice that the integrals of motion (\ref{csymmet10}) are of $(m+n)$th-order in the momenta.  However, since $X^{\pm}$ are, in fact,  complex functions we   can obtain  two  real constants of motion   by  considering their real and imaginary parts, namely, 
\be
\xx= \frac 12(X^+ + X^-),\qquad \yy= \frac 1{2i} (X^+ - X^-),
\label{aj}
\ee
whose maximal order in the momenta is at the most equal to $(m+n)$.   Alternatively, the modulus and the phase functions of  (\ref{csymmet10}) could be considered (as, e.g., in~\cite{Stefan}). 

 In this way we recover the known results on the (super)integrability of anisotropic oscillators on the Euclidean plane~\cite{Jauch, Stefan, Winternitz}:

\vfil\eject
\medskip

\noindent
{\bf Theorem 1.} {\em {\rm (i)} The Hamiltonian $H$ (\ref{ae}) is integrable for any value of $\gamma$, since it is endowed with the quadratic constant of motion given by   $H^\xi$ (\ref{af}).

\noindent
 {\rm (ii)} When $\gamma=m/n$ is a rational parameter (\ref{ah}) the Hamiltonian (\ref{ae}) defines a  superintegrable  anisotropic oscillator with commensurate frequencies $\omega_x:\omega_y$, and the additional real constant of motion  is given by (\ref{aj}), which is at most of $(m+n)$th order in the momenta. The   set $(H,H^\xi,\xx)$ (or $(H,H^\xi,\yy)$) is formed by three functionally independent integrals.
}

\medskip

Some comments concerning the specific anisotropic Euclidean oscillators comprised by the Hamiltonian $H$ (\ref{ae}) seem to be pertinent. 

\begin{itemize}

\item The isotropic  or   1\,:\,1  oscillator corresponds to    $\gamma= m=n=1$, $\omega_x=\omega_y=\omega$,  $\xi=x$ and $p_\xi=p_x $.  
In this case,  the function $\xx$   (\ref{aj}) is a  quadratic integral,  corresponding to    one of the components of the Demkov--Fradkin tensor~\cite{Demkov,Fradkin}, meanwhile $\yy$ (\ref{aj}) is a linear integral in the momenta which is proportional to the angular momentum.

\item The   2\,:\,1  oscillator comes out by setting   $\gamma=2$, $m=2$,  $n=1$,  $\omega_x=2\omega_y=2\omega$, $\xi=2 x$ and $p_\xi =p_x/2$. 
In this case, $\xx$  is a  quadratic integral, meanwhile    $\yy$ is a cubic one.  Thus, the $2:1$ oscillator is considered as a superintegrable system with   quadratic  constants of motion. Obviously, the 1\,:\,2 ($\gamma=1/2$)  oscillator is a completely equivalent system to the   2\,:\,1 ($\gamma=2$) oscillator.

\item 
We remark that    the $1:1$ and $2:1$ (or $1:2$) oscillators are the only anisotropic oscillators endowed with quadratic integrals   according to  the classification of superintegrable systems on the two-dimensional Euclidean space~\cite{mariano99} (see~\cite{evans} for    three dimensions). In other words, all the remaining $m:n$ oscillators have higher-order integrals.

\end{itemize}

It is also worth stressing that for the study of anisotropic  Euclidean oscillators it is not necessary to introduce neither the parameter $\gamma$ (\ref{ab}) nor the new variable $\xi$  (\ref{ab2}) (see the procedure developed in~\cite{Jauch, Stefan, Winternitz}). Nevertheless, such parameter and variable turn out to be essential for defining anisotropic oscillators on spaces of constant curvature which, to the best of our knowledge, were so far unknown.
In the next section we solve this problem, for the first time,  on the two-dimensional sphere.

\section{Anisotropic   oscillators on the sphere}

Let us consider the two-dimensional sphere $ {\mathbf S}^2$ with unit radius embedded in the three-dimensional space  $\mathbb R^{3}$ with ambient coordinates 
$(x_0,x_1,x_2)$ such that
$$
  x_0^2+ x_1^2+   x_2^2   =1 .
$$
We set the origin $O$ in $ {\mathbf S}^2$ as the point given by $O=(1,0,0)\in \mathbb R^{3}$ and we parametrize the ambient coordinates in terms of two intrinsic quantities $(r, \te)$ and  $(x,y)$    in the form
\bea
&& x_0=\cos r=\cos x \cos y  ,\nonumber\\[2pt]
&& x_1=\sin r\cos\te= \sin x \cos y   , \nonumber\\[2pt]
&& x_2=\sin r\sin\te= \sin y   ,
\label{cf}
\eea
whose     geometrical meaning    is as follows~\cite{mariano99,conf, Letter}.

Let $l_1$, $l_2$ be   two base geodesics orthogonal at $O$ and let  $l$ be the geodesic  that joins a point $P$ (the particle) and the origin $O$. The so called  {\em geodesic polar coordinates}  $(r,\phi)$ are defined by the   distance $r$  between $O$ and $P$ measured  along $l$ and the   angle $\phi$ of $l$ relative to $l_1$. Thus these generalize  the    polar coordinates to the sphere. Now let $P_1$ be the intersection point of $l_1$ with its orthogonal geodesic $l'$ through $P$. Then  the so called  {\em geodesic parallel coordinates}  $(x,y)$ are determined by  the distance $x$  between $O$ and the point  $P_1$  measured  along $l_1$    and  the distance $y$  between $P_1$ and $P$ measured  along $l'$ (see Figure~\ref{g1}).   Hence these generalize  the    Cartesian coordinates to $ {\mathbf S}^2$. The domain of these variables reads
$$
  0< r<  {\pi}  ,\qquad  0 \le \phi<2 \pi  ,\qquad     - {\pi} < x\le  {\pi} ,\qquad   -\frac{\pi}{2}< y< \frac{\pi}{2} .      \label{zb}
$$

  \begin{figure}
\includegraphics[width=0.4\textwidth]{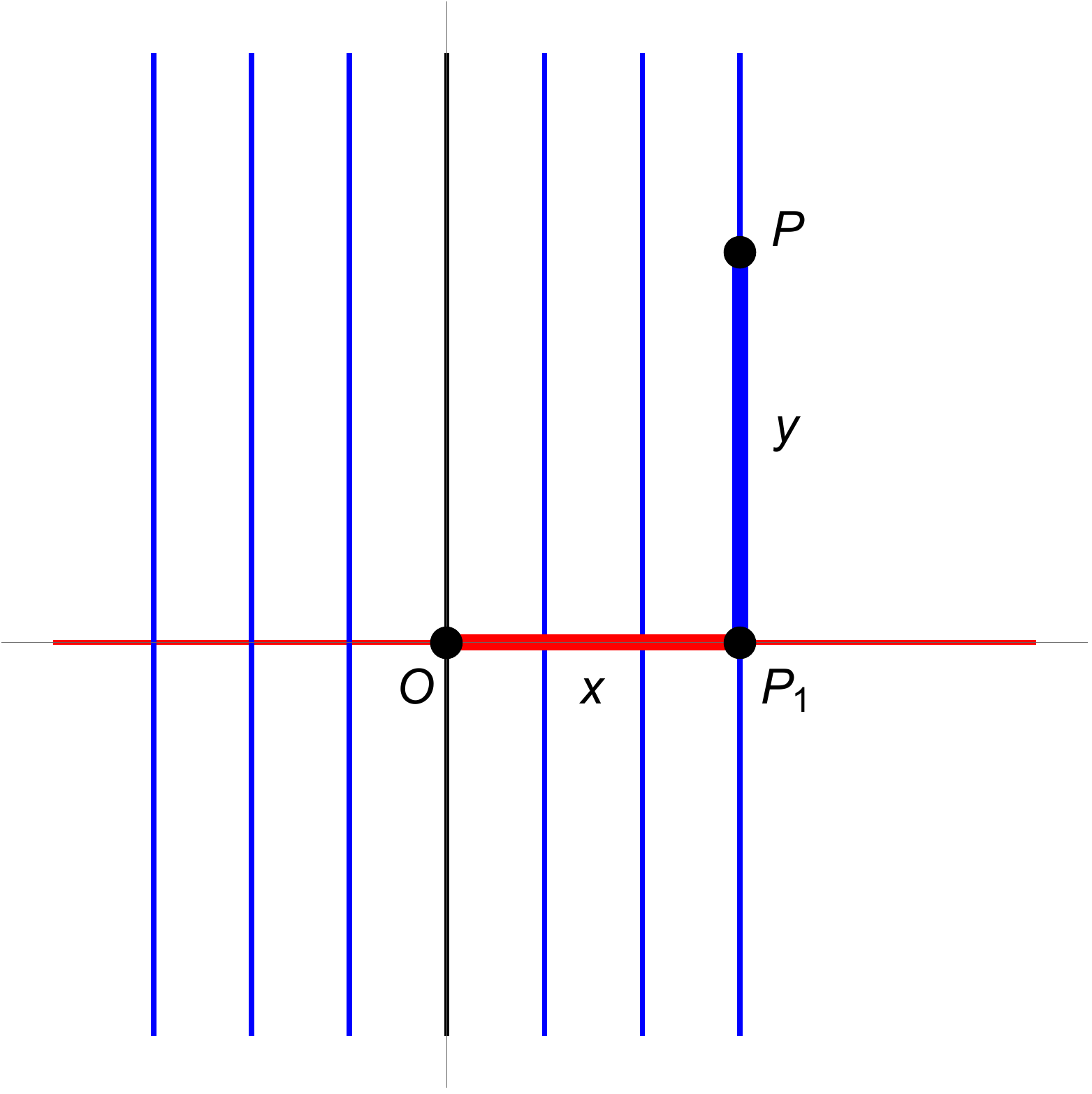}
\qquad
\includegraphics[width=0.4\textwidth]{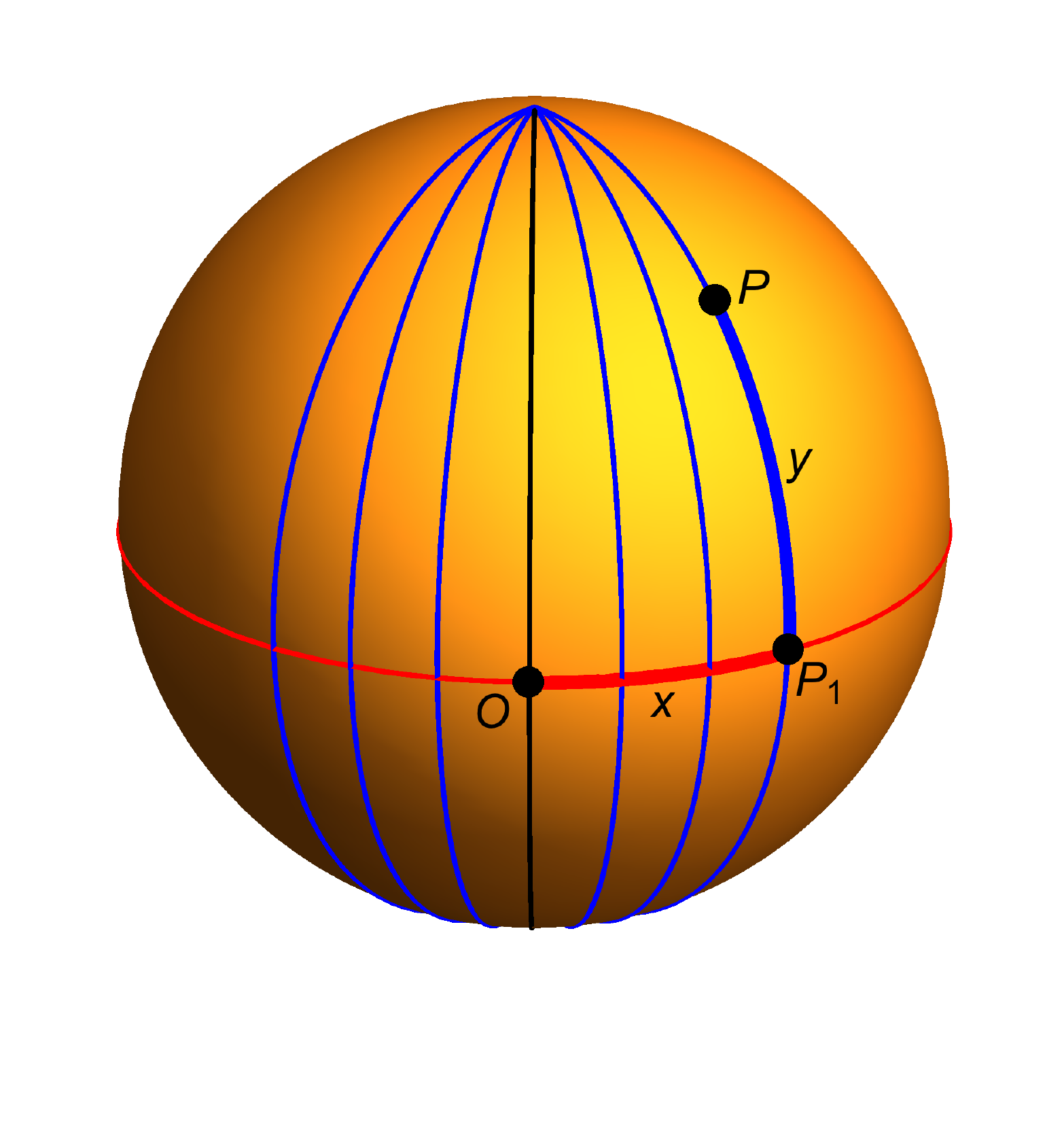}
\caption{Cartesian coordinates in the plane versus geodesic parallel coordinates on the sphere.}
  \label{g1}
\end{figure}

 The   metric on  $ {\mathbf S}^2$ in the above coordinates  is so given by:
$$
{\rm d} s^2= \left(\left. {\rm d} x_0^2+    {\rm d} x_1^2+   {\rm d} x_2^2 \right) \right|_{{\mathbf S}^2}=      {\rm d} r^2+  \sin^2 r \, {\rm d} \phi^2   =\cos^2 y\, {\rm d} x^2 + {\rm d} y^2  .
$$
By denoting  $(p_r,p_\te)$  and    $(p_x,p_y)$  the conjugate momenta of $(r,\te)$ and $(x,y)$, respectively, we obtain the free Hamiltonian $T$ that determines the free motion  on $ {\mathbf S}^2$:
$$
 {T}=\frac 12 \left( p_r^2+\frac{p_\te^2}{\sin^2 r} \right) =\frac 12 \left(\frac{p_x^2}{\cos^2  y}  +  p_y^2\right)  .
 \label{ci}
$$

Now, by having in mind the Euclidean Hamiltonian (\ref{ae}), we are able to propose an ``appropriate"  Hamiltonian that determines anisotropic oscillators on the sphere. Explicitly, we shall consider  the following Hamiltonian,  expressed  in terms of the geodesic parallel variables, and given  by
\begin{equation}\label{hc1}
{  H} =  {  T}+ {  U}^\gamma = \frac{1}{2} \left(\frac{p_{x}^2}{ \cos  ^2y}+p_{y}^2\right)+\frac{\omega^2}{2} 
\left(\frac{  \tan ^2(\gamma x)}{ \cos^2y}+\tan^2y\right) . 
\end{equation}
  We remark that due to  term  $\tan(\gamma x)$ in the potential,  the domain of the variable $x$ and the value of the real parameter $\gamma$ are restricted in the form
 $$
    -\frac{\pi}{2 }< \gamma x< \frac{\pi}{2 } ,\qquad      \gamma\ge \frac 12  ,
  \label{range}
$$
 so avoiding a multivalued Hamiltonian.

Next  we introduce the variable $\xi$ (\ref{ab2}) and we write the Hamiltonian (\ref{hc1}) as 
 \bea
 \label{hc2}
&& {  H} = 
\frac{p_{y}^2}{2}+\frac{1}{\cos^2 y }\left(\frac{p_{x}^2}{2}+
\frac{\omega^2}{2  \cos^2(\gamma x)}\right)-
\frac{\omega^2}{2 }  \nonumber\\[2pt]
&& \quad\, = \frac{p_{y}^2}{2}+\frac{\gamma^2}{ \cos^2 y }\left(\frac{p_{\xi}^2}{2}+\frac{\omega^2}{2    \gamma^2 \cos^2\xi}\right)-
\frac{\omega^2}{2   } ,
\label{hc3}
\eea
 which leads to a quadratic integral of motion $H^\xi$ such that 
\begin{equation}\label{hc31}
H   = \frac{p_{y}^2}{2}+\frac{\gamma^2 H^\xi}{\cos^2y}-
\frac{\omega^2}{2 } , \qquad   H^\xi = \frac{p_{\xi}^2}{2}
+\frac{\omega^2}{2 \gamma^2\cos^2\xi} ,\qquad \{ H  , H^\xi \}=0 .
\end{equation}
Hence, when $H^\xi $ is taken as a constant, $H$ is reduced to a   one-dimensional Hamiltonian   which
 always determines an integrable  anisotropic    oscillator on $ {\mathbf S}^2$   for any value of $\gamma$.  
 
In the sequel we will factorize  the one-dimensional Hamiltonians $H^\xi$  and  $H$ (\ref{hc31}).  The  resulting (ladder and shift)   factor functions  will  provide additional integrals of motion of the two-dimensional Hamiltonian 
(\ref{hc3}) whenever $\gamma$ is a rational number, similarly to what happens for the previous  anisotropic Euclidean oscillators.

\subsection{Factorization}

 Let us consider the  one-dimensional integral of motion $H^\xi$ (\ref{hc31}) and look for  some ladder functions  $B^\pm$   fulfilling  the Poisson brackets
\be
\{H^\xi , B^\pm\}=f_\pm(H^\xi) B^\pm ,
\label{poisB}
\ee
for certain functions $f_\pm$.  We define the following function 
$$
h^{\xi} =\cos^2\xi\left(\frac{p_{\xi}^2}{2}-H^\xi \right).
$$
As in (\ref{qa}),  we require that
$$
h^{\xi} =B^+B^-+\lambda_{B}  ,
$$
which yields, as particular solutions, to
\begin{equation}\label{bc}
B^{\pm} ={\mp}\frac{i}{\sqrt{2}}\cos \xi\, p_{\xi}+
\sqrt{  H^{\xi}}\sin \xi,\qquad  \lambda_{B}=
- H^{\xi}  .
\end{equation}
Although is clear that $h^{\xi} \equiv - {\omega^2}/({2 \gamma^2})$, the remarkable point is that $B^{\pm}$ are ladder functions for  $H^\xi$ (\ref{hc31}) since  the three  functions  $(H^\xi,B^\pm)$  verify the Poisson brackets
\begin{equation}\label{commpt1}
\{H^\xi,B^{\pm}\}= \mp i\sqrt{{2  H^\xi}}\,B^{\pm} ,\qquad \{ B^-,B^+\}=- i  \sqrt{{2  H^\xi}} ,
\end{equation}
which are of the type (\ref{poisB}). Notice that we have thus obtained a ``deformation" of the  Poisson--Lie   algebra $\mathfrak{h}_4$   (\ref{commpt20}).

The expressions (\ref{commpt1})  suggest to   write the integral  $H^\xi$ (\ref{hc31}) as
\be
\mm= \sqrt{2  H^\xi} .
\label{mm}
\ee
The Poisson brackets among the  Hamiltonian ${H}$ (\ref{hc31}) and the ladder functions  $B^{\pm}$ (\ref{bc})  read 
$$
\{H ,B^{\pm}\}=\mp  i\,\frac{\gamma^2 \mm}{\cos^2 y}\,B^{\pm},\qquad \{ B^-,B^+\}=- i \, \mm .
$$

The second step in the factorization approach is to search for   shift  functions $A^\pm$ for the Hamiltonian $H$  (\ref{hc31}), written in terms of (\ref{mm}) as 
$$
H   = \frac{p_{y}^2}{2}+\frac{\gamma^2 \mm^2}{2\cos^2y}-
\frac{\omega^2}{2 }, 
$$
which must verify the Poisson brackets
$$
\{H  , A^\pm\}=g_\pm(H^\xi,U^\gamma) A^\pm =g_\pm(\mm,y) A^\pm ,
$$
for certain functions $g_\pm$ and where $U^\gamma$ is the potential given in (\ref{hc1}). We impose the relation (\ref{qb}), finding now  that
\begin{equation}\label{capm}
A^{\pm}=\mp \frac{i}{\sqrt{2}}\,p_{y}-\frac{\gamma \mm}{\sqrt{2} }\tan y,\qquad 
\lambda_{A} =\frac{1}{2 }  \left(\gamma^2\mm^2-\omega^2\right) .
\end{equation}
From this result we find the following Poisson brackets
$$
\{H ,A^{\pm}\}=\pm i\,\frac{\gamma\mm}{\cos^2y}\,A^{\pm} ,\qquad \{ A^-,A^+\}=i\gamma\,\frac{\mm}{\cos^2 y} .
$$

\subsection{Higher-order integrals of motion}

Similarly to the  anisotropic oscillators on the Euclidean plane, it is a matter of straightforward computations to show that  for rational values $\gamma=m/n$ (\ref{ah}), we obtain (higher-order) additional integrals of motion for   $H$ (\ref{hc1}), namely
\bea
&& X^{\pm}=(B^{\pm})^n (A^{\pm})^{m} ,\qquad    \{ H, X^{\pm}\}=0, \nonumber\\[2pt]
&&  \xx= \frac 12(X^+ + X^-),\qquad \yy= \frac 1{2i} (X^+ - X^-),
\label{csymmet1}
\eea 
where $B^{\pm}$ and $A^{\pm}$ given in (\ref{bc}) and (\ref{capm}), respectively.

Consequently, the  generalization of  Theorem 1 on the sphere   is obtained, and a new infinite family of (super)integrable curved systems are found:
\medskip

\vfil\eject

\noindent
{\bf Theorem 2.} {\em {\rm (i)} The Hamiltonian $H$  (\ref{hc1}) defines an   integrable anisotropic   oscillator on ${\mathbf S}^2$   for any value of the positive real parameter $\gamma$. The    (quadratic in the momenta)  constant of motion for $H$ is given by   $H^\xi$   (\ref{hc31}).

\noindent
 {\rm (ii)} When $\gamma$ is a rational parameter (\ref{ah}),   the Hamiltonian  $H$  (\ref{hc1})  provides   a  superintegrable  anisotropic   oscillator  on ${\mathbf S}^2$ with    additional constants of motion    given by (\ref{csymmet1}), which are at most of $(m+n)$th  order in the momenta. The   set $(H,H^\xi,\xx)$ (also $(H,H^\xi,\yy)$) is formed by three functionally independent integrals.
}

\medskip   

We stress that this statement covers two well-known particular  cases which correspond  to  $\gamma=1$ and $\gamma=2$,  respectively. Indeed, these are the two cases appearing 
 in   the classification of quadratic superintegrable systems  on the sphere~\cite{mariano99,Kalnins1,Kalnins2}, meanwhile all the remaining ones are new superintegrable systems  on ${\mathbf S}^2$. 

In particular, the curved oscillator system coming from $\gamma=m=n=1$, $\xi=x$ and $p_\xi=p_x $ is just  the
 so called   Higgs oscillator~\cite{Higgs, Leemon}  whose potential, expressed  in geodesic polar coordinates (\ref{cf}),  is  simply  $\tan^2 r$. This curved oscillator    has been widely studied (see, e.g., ~\cite{
Letter, Pogoa, Nersessian1, Santander2, Santander3, Santander4, Santander4b, ballesteros, Santander6, Nersessian2, Annals09, ballesteros13} and references therein).
In both sets of coordinates (\ref{cf}), the Higgs potential  reads 
$$
  {  U}^{\gamma=1} =  \frac{\omega^2}{2} 
\left(\frac{  \tan ^2  x}{ \cos^2y}+\tan^2y\right) =\frac{\omega^2}2  \tan^2 r. 
$$

The case  with  $\gamma=2$ was firstly introduced in the classification presented in~\cite{mariano99}  (see also~\cite{ballesteros13, anisotropic}). In  geodesic parallel and polar coordinates (\ref{cf}) this potential is given by
\bea
&& U^{\gamma=2} = \frac{\omega^2}{2} \left( \frac{ \tan^2(2x) }{  \cos^2 y } +   \tan^2 y  \right)  \nonumber\\[2pt]
&& \qquad \  = \frac{\omega^2}{2} \left( \frac{4\tan^2  r\cos^2\te}{\left(1-  \sin^2 r\sin^2\te\right)\left(1-  \tan^2 r \cos^2\te\right)^2} +  \frac{\sin^2 r\sin^2\te}{ 1-  \sin^2 r\sin^2\te }     \right)  .
\nonumber
\eea
The latter expression clearly justifies the use of geodesic parallel variables instead of the geodesic polar ones when looking for generic anisotropic oscillators on   ${\mathbf S}^2$. We also remark that,  in contradistinction with the anisotropic oscillators on the  Euclidean plane, the curved potentials 
 $U^{\gamma} $ and $U^{1/\gamma} $ are no longer equivalent.


  
\section{The TTW system on the Euclidean plane}

In this section we will consider another example of superintegrable system on
the Euclidean plane, but using polar coordinates instead of Cartesian ones. This is the case
of the well-known TTW system \cite{ttw09}. In what follows we will apply 
the factorization method to find the symmetries according to 
\cite{Kuru3}, although we will introduce some inessential changes in order to simplify the discussion.

The TTW Hamiltonian, in   polar coordinates $(r,\phi)$, 
has the following expression (the factor $1/2$ has been suppressed to
accomodate with the notation of \cite{Kuru3}):
\be
H \displaystyle = p^2_r +\omega^2 r^2 + \frac{1}{r^2}
\left(p_\phi^2 + \frac{\gamma^2\alpha^2}{\cos^2 (\gamma\phi)}
+\frac{\gamma^2\beta^2}{\sin^2 (\gamma\phi)}\right)  .
\label{TTWa}
\ee
Here it is assumed that $\gamma\geq 1/4$, in such a manner that  the potential is well defined for 
$0<\gamma \phi<\pi/2$.
If we perform the usual canonical transformation
$\theta= \gamma \phi$,  $p_\theta = p_\phi/\gamma$, 
then the TTW Hamiltonian becomes
\begin{equation}\label{hcm}
H=p^2_r +\omega^2 r^2 + \frac{\gamma^2 H_\theta}{r^2} ,\qquad H_\theta = p^2_\theta + \frac{\alpha^2}{\cos^2\theta}
+\frac{\beta^2}{\sin^2 \theta} ,
\end{equation}
yielding two one-dimensional systems, that is, the angular Hamiltonian $H_\theta$ and the radial one $H\equiv H_r$, provided that   $H_\theta$ is assumed to take a fixed constant value.
 
In this case we shall also deal with  ``ladder" and ``shift" functions separately for the angular and radial parts of the TTW Hamiltonian function (\ref{hcm}), and the symmetries will be constructed in the 
same way as in the previous examples. 

\subsection{Factorization}

We recall that 
the Hamiltonian $H_\theta$   (\ref{hcm}) is known as the 
two-parameter P\"oschl-Teller Hamiltonian \cite{calzada12}.
The ladder functions
have a similar expression as in the quantum case~\cite{calzada12}, namely
\begin{equation}\label{bc2}
B^{\pm} ={\pm}i\sin 2 \theta\,  p_\theta+\sqrt{H_\theta} \cos 2
\theta+\frac{\beta^2-\alpha^2}{\sqrt{H_\theta}}\, .
\end{equation}
Along with the Hamiltonian  $H_\theta$, they satisfy the following Poisson brackets 
\begin{equation}\label{hbb}
\{H_\theta,B^{\pm}\}= \mp 4 i\sqrt{{H_\theta}}\,B^{\pm} , \qquad \{ B^-,B^+ \} = -4 i \sqrt{{H_\theta}}\left(1-\frac{(\beta^2-\alpha^2)^2}{H_\theta^2} \right).
\end{equation}
The product of these two functions gives another function 
which depends only on  $H_\theta$:
$$
B^{+}B^{-}=H_\theta+\frac{(\beta^2-\alpha^2)^2}{H_\theta}-2(\beta^2+\alpha^2) .
$$
 
The second Hamiltonian involved in (\ref{hcm}) is the radial oscillator Hamiltonian:
$$
H=H_{r} = p_r^2 +\omega^2 r^2 +\frac{\gamma^2 \mpt^2}{r^2} ,\qquad  \mpt=\sqrt{H_\theta} .
$$
This is factorized in a similar way to the quantum case
\cite{david} by requiring the relation  (\ref{qb}) which yields   two types of shift functions:
\bea
&& A_1^{\pm} = \mp i  p_r + \omega  r - \frac{\gamma \mpt}{r}, \qquad \lambda_{1A}=
2 \omega \gamma \mpt , \nonumber\\[2pt]
&&A_2^{\pm} = \mp i  p_r + \omega  r + \frac{\gamma \mpt}{r}, \qquad \lambda_{2A}=
-2 \omega \gamma \mpt \,.
\nonumber
\eea
These functions together with $H_{r}$ satisfy the following
 Poisson brackets
\bea
&&\{H_r,A_1^{\pm}\}= \mp 2 i \left(\omega+\frac{\gamma \mpt}{r^2}\right)  A_1^{\pm}, \qquad \{A_1^{-},A_1^{+}\}=  -2i\left( \omega +\frac{\gamma \mpt}{r^2}\right),\nonumber\\
&&\{H_r,A_2^{\pm}\}= \mp 2 i \left(\omega-\frac{\gamma \mpt}{r^2} \right) A_2^{\pm} , \qquad \{A_2^{-},A_2^{+}\}=  -2i\left( \omega -\frac{\gamma \mpt}{r^2}\right)  .
\nonumber
\eea
They are  ``mixed" ladder-shift functions~\cite{Kuru3,calzada12}, but
we can construct ``pure" shift functions by taking the following products:
\begin{equation}\label{aes}
A^+ = A_1^+A_2^-,\qquad A^- = A_1^-A_2^+ ,
\end{equation}
satisfying
\begin{equation}\label{paes}
\{H_r,A^{\pm}\}= \mp 4 i\, \frac{\gamma \mpt}{r^2}\,A^{\pm} ,\qquad \{A^{-} , A^{+}\}=  - 8i\,\frac{\gamma \mpt }{r^2}\, H_r.
\end{equation}
The above shift functions are different from those  given in \cite{calzada12}, but
they lead to similar results.

\subsection{Higher-order integrals of motion and trajectories}

Whenever $\gamma=m/n$, additional symmetries $X^\pm$ for the Hamiltonian (\ref{hcm}) can be easily obtained in terms of the functions 
 $B^{\pm}$ (\ref{bc2}) and $A^{\pm}$ (\ref{aes}) in the same form given in  (\ref{csymmet1}).    This is proved,  directly, with the help of (\ref{hbb}) and (\ref{paes}).   Consequently, for a rational value of $\gamma$, the TTW Hamiltonian (\ref{TTWa})  determines a superintegrable system.

These symmetries are helpful in order to find quite easily the phase trajectories 
for the Hamiltonian (\ref{hcm}). 
When we fix the value of any of these real symmetry functions (i.e.~$X$ or  $Y$  (\ref{csymmet1}))  together with
$H$ and $H_\theta$, then we get a trajectory.
Some examples for different values of $\gamma$ have been plotted in Figures 2 and 3 in
the $(r,\theta)$-plane. If one preferes to deal with the $(r,\phi)$-plane, it is enough to
change the angle, i.e.~$\theta = \gamma \phi$, but the shape will remain quite similar.

For the discussion concerning the algebra generated by the integrals of motion and the corresponding polynomial symmetries we refer the reader to~\cite{Kuru4}.

\begin{figure}
\begin{center}
\includegraphics[width=0.4\textwidth]{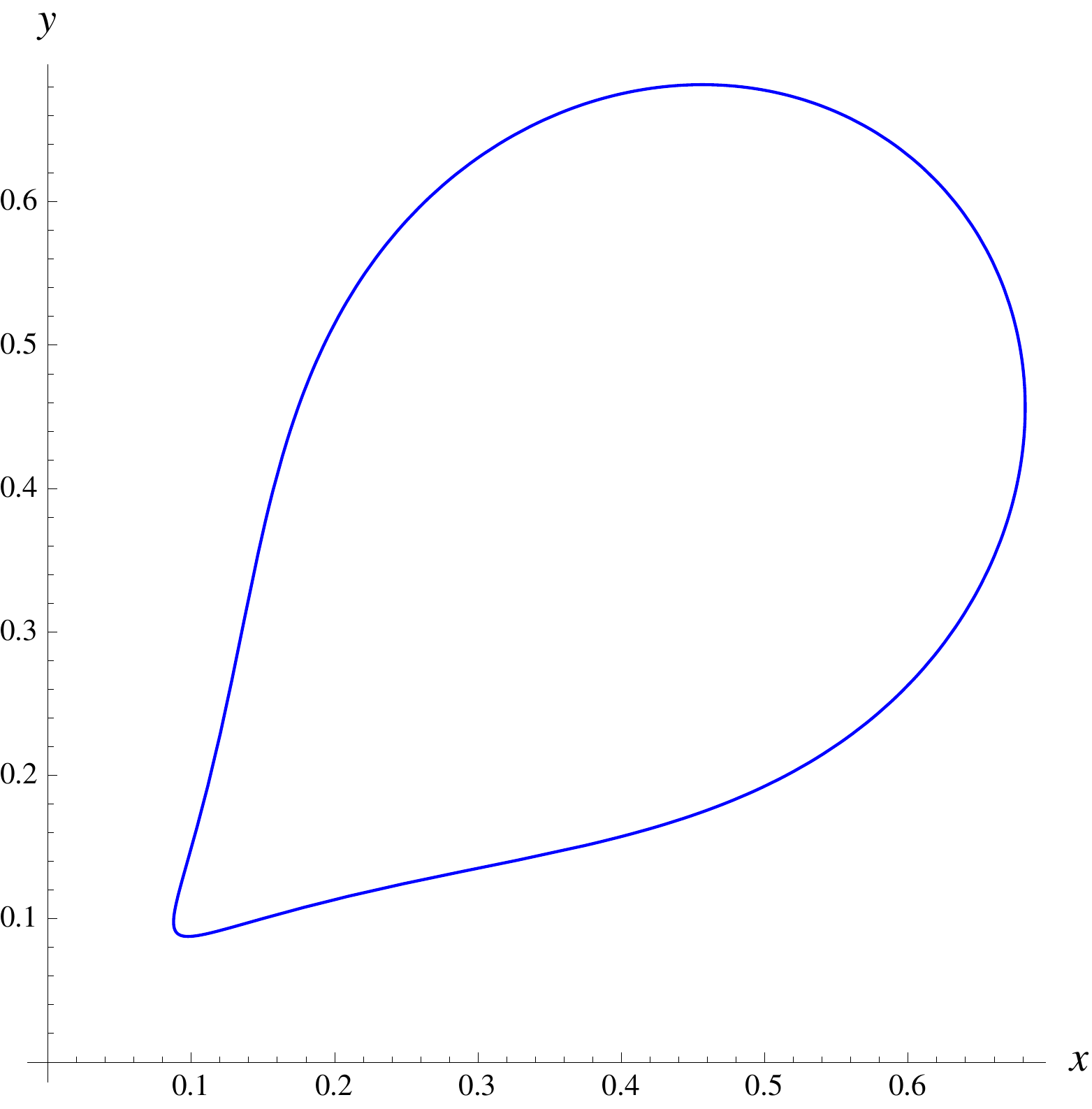}\quad\quad\quad
\includegraphics[width=0.4\textwidth]{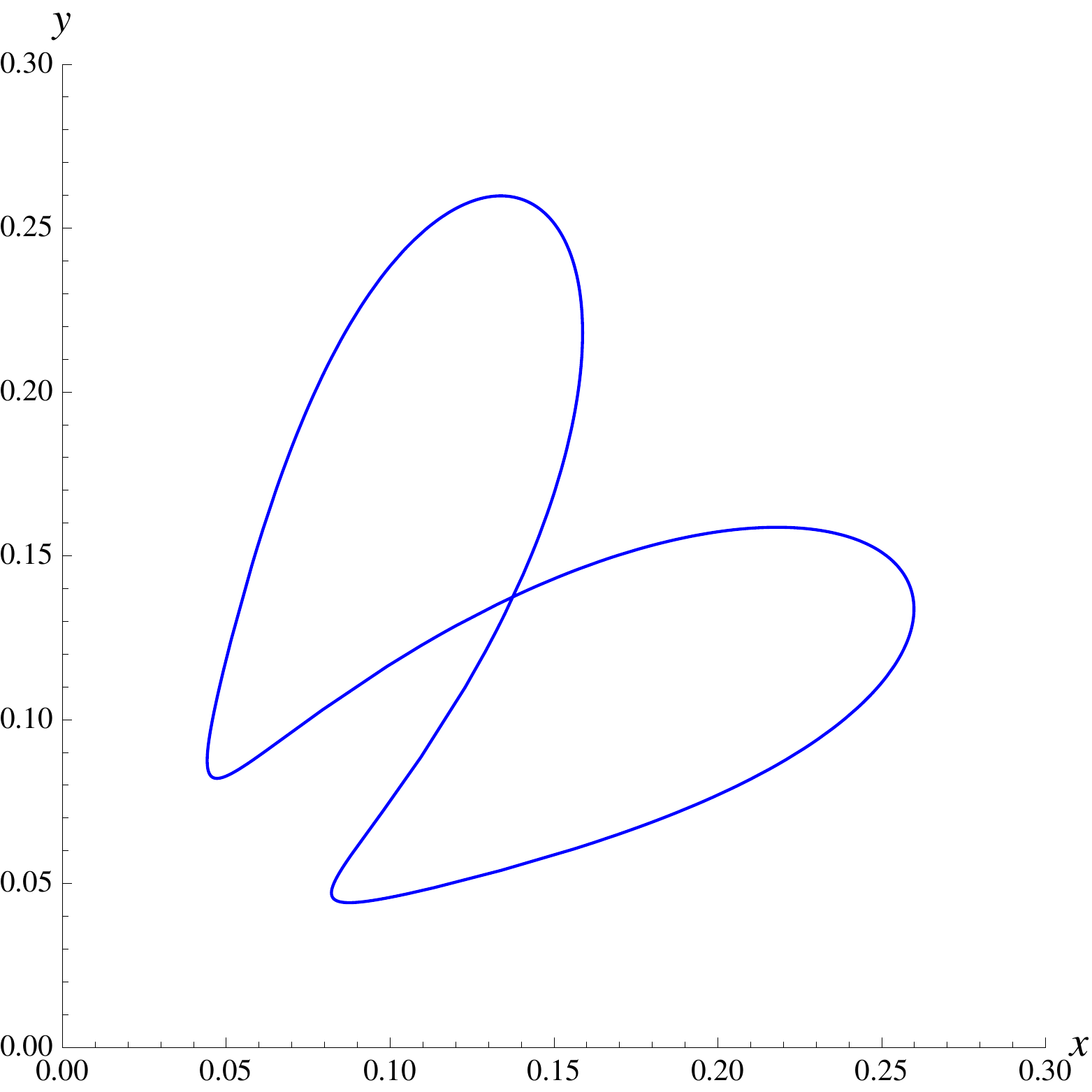}
\caption{Trajectories in the $(r,\theta)$-plane corresponding to: $\gamma=1$ (left) 
and $\gamma=2$ (right).
\label{fig1}}
\end{center}
\end{figure}

\begin{figure}
\begin{center}
\includegraphics[width=0.4\textwidth]{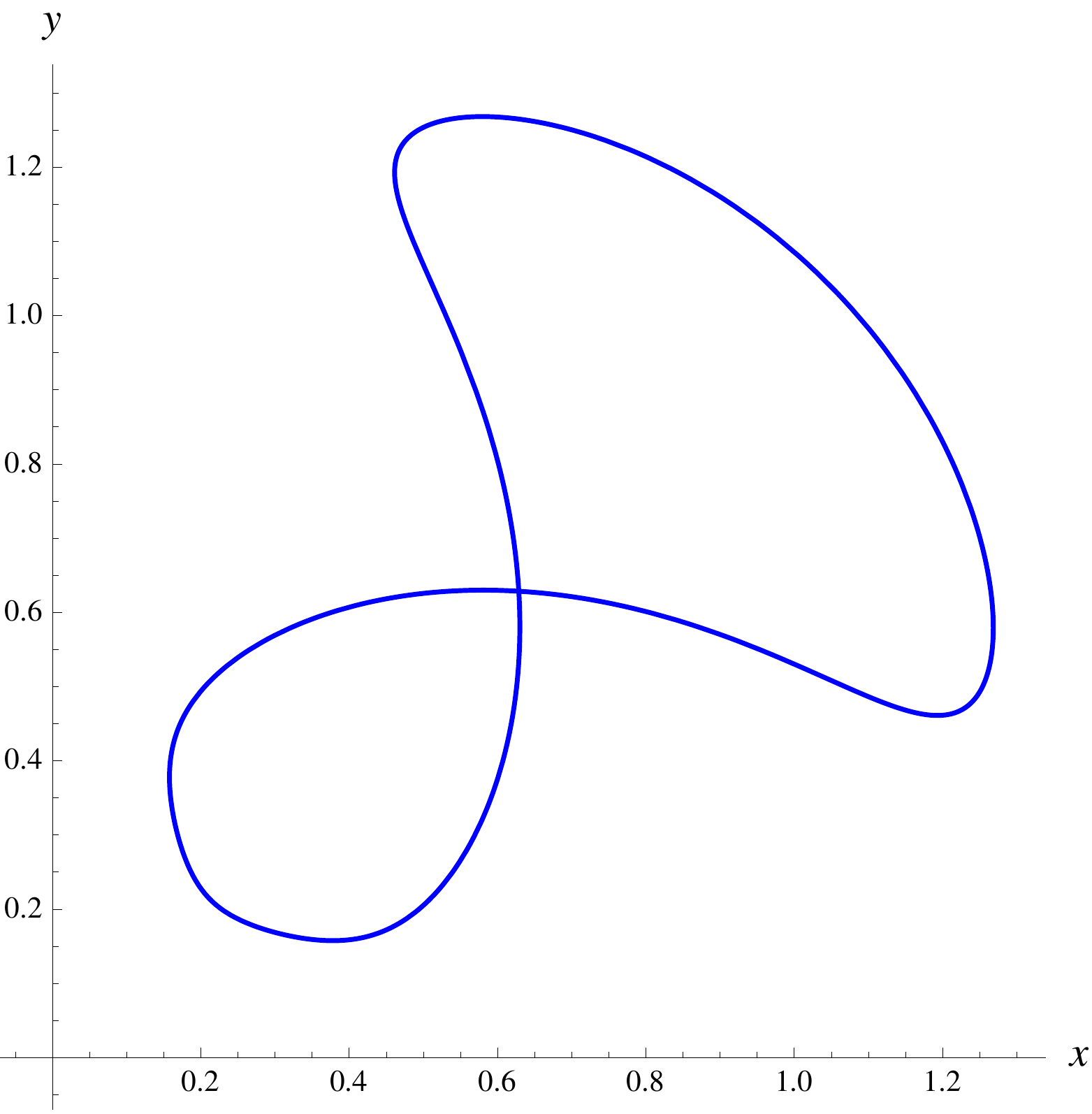}\quad\quad\quad
\includegraphics[width=0.4\textwidth]{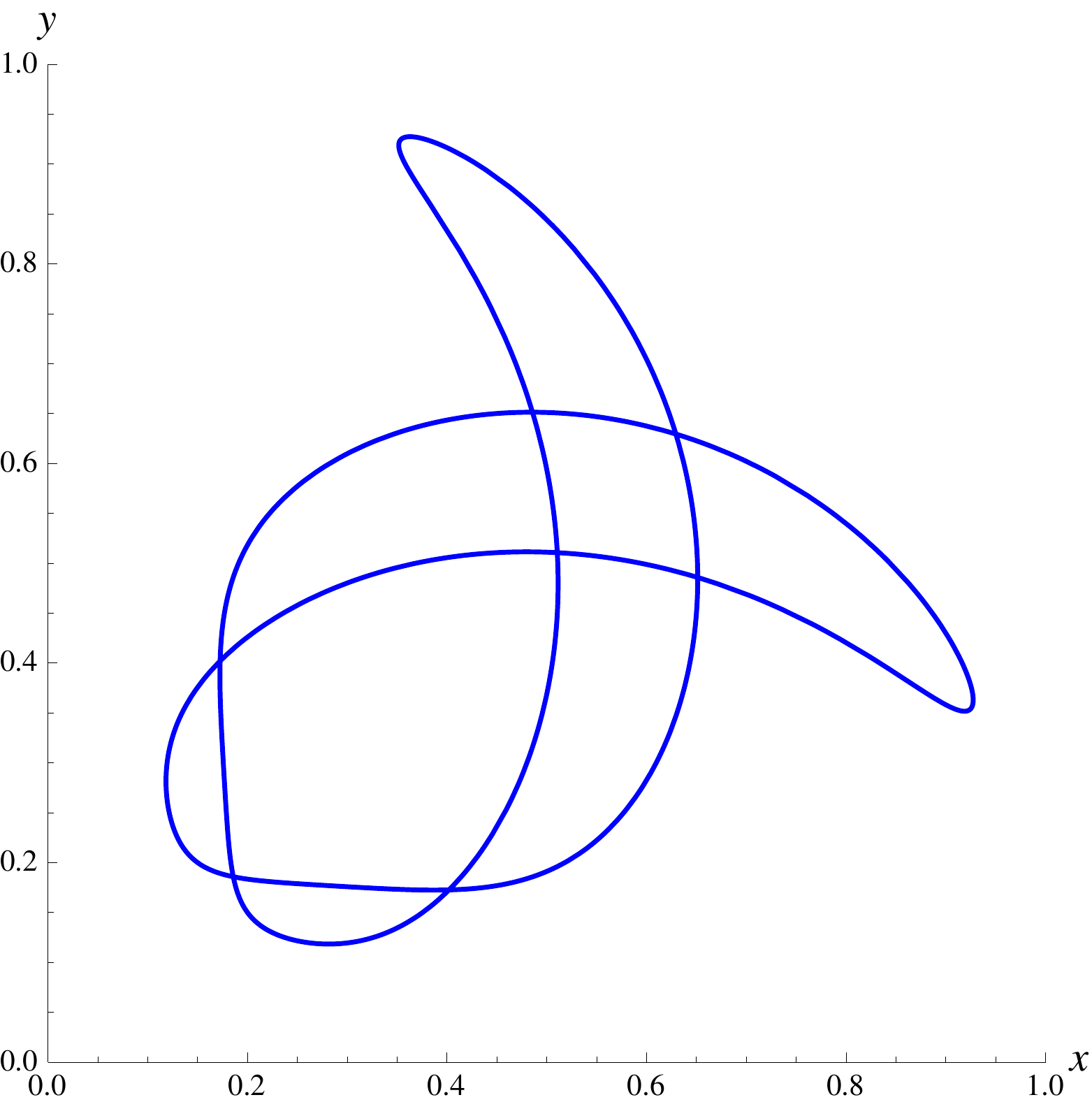}
\caption{Trajectories in the $(r,\theta)$-plane corresponding to: 
$\gamma=1/2$ (left) and $\gamma=2/3$ (right).
\label{fig2}}
\end{center}
\end{figure}

  \newpage

  
\section{Concluding remarks }

The factorization approach to integrable systems has been revisited, and we have illustrated it by considering known integrable Hamiltonians, such as the anisotropic oscillator and   the TTW  system on the Euclidean plane. Also, new systems like the anisotropic oscillator on the sphere have been introduced with the aid of this technique. It would be indeed interesting to apply this approach to other relevant problems, for instance:
\begin{itemize}

\item To construct curved anisotropic oscillators on other spaces of constant curvature such as the hyperbolic space as well as onto the relativistic  (anti-)de Sitter  and Minkowski spacetimes (see~\cite{ballesteros} for the $1:1$ oscillator on these spacetimes and, more recently,~\cite{so221, so22} for the   oscillator  problem on the $SO(2,2)$ hyperboloid).

\item The addition of  two ``centrifugal potentials", that on the sphere can be considered as noncentral oscillators, by keeping superintegrability~\cite{Letter, ballesteros13}. We recall that in the Euclidean space the superposition of the anisotropic oscillator  with centrifugal terms leads to the so called caged anisotropic oscillator, studied in~\cite{Verrier}.

\item  To  study superintegrable Hamiltonian systems  on spaces of nonconstant curvature. Recently, examples of this type of systems have been presented in~\cite{Ragnisco1} (a curved Kepler--Coulomb problem on the  Taub-NUT space~\cite{taub}) and in~\cite{Ragnisco2}  (a curved oscillator on the Darboux III space~\cite{darboux}). 

\item And, finally and more importantly, to define and solve the corresponding quantum systems. 

\end{itemize}

Work on all these lines is currently in progress and will be presented elsewhere.


\section*{{Acknowledgments}}

This work was partially supported by the Spanish Ministerio de Econom\1a y Competitividad     (MINECO) under grants   MTM2013-43820-P and MTM2014-57129, and by the Spanish Junta de Castilla y Le\'on  under grant BU278U14.     \c{S}.~Kuru acknowledges the warm hospitality at the  Department of
Theoretical Physics, University of Valladolid, Spain.


\end{document}